%% file: 0-paper.tex
\documentclass[a4paper,UKenglish,cleveref, autoref, thm-restate]{lipics-v2021}

\pdfoutput=1 
\hideLIPIcs 

\usepackage{makecell}
\usepackage{float}
\usepackage{listings}
\usepackage{longtable}
\usepackage{booktabs}

\title{Much of Geospatial Web Search Is Beyond Traditional GIS}

\author{Ilya Ilyankou\footnote{Corresponding author}}{SpaceTimeLab, Department of Civil, Environmental, and Geomatic Engineering, UCL, London, UK \and \url{https://ilyankou.com} }{ilya.ilyankou.23@ucl.ac.uk}{https://orcid.org/0009-0008-7082-7122}{This work was supported by Ordnance Survey \& UKRI Engineering and Physical Sciences Research Council [grant no. EP/Y528651/1].} 

\author{Stefano Cavazzi}{Ordnance Survey, Southampton, UK}{stefano.cavazzi@os.uk}{https://orcid.org/0000-0003-3575-0365}{}

\author{James Haworth}{SpaceTimeLab, Department of Civil, Environmental, and Geomatic Engineering, UCL, London, UK}{j.haworth@ucl.ac.uk}{https://orcid.org/0000-0001-9506-4266}{}

\authorrunning{I. Ilyankou, S. Cavazzi, and J. Haworth} 

\Copyright{Ilya Ilyankou, Stefano Cavazzi, and James Haworth} 

\ccsdesc[500]{Information systems~Web searching and information discovery}
\ccsdesc[500]{Information systems~Geographic information systems}
\ccsdesc[500]{Information systems~Web log analysis}
\ccsdesc[300]{Computing methodologies~Natural language processing}
\ccsdesc[300]{Computing methodologies~Cluster analysis}
\ccsdesc[300]{Human-centered computing~Human computer interaction (HCI)}

\keywords{Web search queries, geographic information retrieval, query classification, geospatial query taxonomy, MS MARCO, sentence embeddings, density-based clustering, GeoAI, large language models, place theory} 

\relatedversion{}
\relatedversiondetails{Full Version}{https://arxiv.org/abs/2605.11336}

\supplement{}
\supplementdetails[subcategory={Code}, cite={}, swhid={}]{Software}{https://github.com/ilyankou/ms-marco-geospatial}
\supplementdetails[subcategory={Geospatial query classifier}, cite={}, swhid={}]{Model}{https://huggingface.co/ilyankou/is-geospatial-query}

\nolinenumbers 

\EventEditors{Sabine Timpf, Gabriele Filomena, Armand Kapaj, Rui Zhu, Nicholas Giudice, and Ed Manley}
\EventNoEds{6}
\EventLongTitle{17th International Conference on Spatial Information Theory (COSIT 2026)}
\EventShortTitle{COSIT 2026}
\EventAcronym{COSIT}
\EventYear{2026}
\EventDate{September 22--25, 2026}
\EventLocation{York, UK}
\EventLogo{}
\SeriesVolume{393}
\ArticleNo{4}

\begin{document}

\maketitle

\begin{abstract}
Web search queries concern place far more often than existing labelling schemes suggest, yet the landscape of geospatial web search queries -- what people ask of place, and how often -- remains poorly characterised at scale. We apply dense sentence embeddings, a lightweight SetFit classifier, and density-based clustering to the full MS MARCO corpus of 1.01 million real Bing queries without prior filtering for toponyms or spatial keywords, identifying 181,827 geospatial queries (18.0\%), nearly threefold the 6.17\% labelled as \textit{Location} in the original annotations. The resulting taxonomy of 88 query categories reveals that geospatial web search is dominated by transactional and practical lookups: costs and prices alone account for 15.3\% of geospatial queries, nearly twice the size of the entire physical geography theme. Much of this activity -- costs, opening hours, contact details, weather, travel recommendations -- falls outside the scope of what traditional GIS and knowledge graphs are built to serve. The categories vary substantially in the kind of answer they admit, from deterministic lookups answerable from spatial databases or knowledge graphs to evaluative or temporally volatile queries that require generative or real-time systems. We discuss implications for hybrid retrieval architectures and for benchmarks of geographic reasoning in large language models. We openly release the labelled dataset, classifier, and taxonomy.
\end{abstract}

\input{1-intro}

\input{2-related-work}

\input{3-methodology}

\input{4-results}

\input{5-discussion}

\input{6-conclusion}



\bibliography{references,manual}

\input{7-appendix}


\end{document}

%% file: 1-intro.tex
\section{Introduction}

Place is central to how people make sense of the world, yet existing information systems represent it in an impoverished way, often reducing it to named points or coordinate-bound objects rather than the vague, relational, and socially constructed phenomenon it actually is \cite{purves_places_2019,cresswell_place_2014,hamzei_place_2020}. Large-scale web search query logs offer an empirical window into what people want from place: whether a town falls in a particular county, what the weather or cost of living is somewhere, where the nearest airport is, or what language is spoken in a country. The MS MARCO corpus of 1.01 million real Bing queries \cite{bajaj_ms_2018} remains the largest publicly available record of such behaviour, despite being collected prior to 2018 and skewed toward Anglophone North American users. It captures web search rather than conversational, voice, or professional GIS use, but it is the most suitable corpus currently available for characterising geospatial web search at scale. 

Prior work has identified geospatial queries primarily by syntactic form or the presence of toponyms \cite{hamzei_initial_2019,hamzei_place_2020,kuhn_semantics_2021}, producing prevalence estimates that are typically too low and taxonomies too narrow to capture the full range of what people want from place. We argue that the right starting point is the query corpus itself, without prior filtering. By applying Transformer sentence embeddings \cite{reimers_sentence-bert_2019} and density-based clustering to the full MS MARCO dataset, we build a data-driven taxonomy of geospatial web search queries that quantifies not just how many queries are geospatial, but what categories they fall into and how those categories are distributed. Much of this activity, such as costs, opening hours, travel recommendations, or weather, falls outside the scope of what traditional GIS is built to serve, with direct implications for hybrid retrieval and generative system design.

We address three research questions:

\begin{enumerate}
    \item What proportion of web search queries are geospatial, under a definition broader than `contains toponyms'?
    \item What are the dominant themes of geospatial web search, and how are they distributed?
    \item How do the resulting categories vary in the kind of answer they admit, and what does this imply for the design of systems that serve such queries?
\end{enumerate}

Our work makes three practical contributions. First, we release a gold human-annotated dataset of 1,200 web search queries, labelled as geospatial or non-geospatial, constructed to cover the full embedding space of MS MARCO rather than its high-density regions. Second, we train and release a lightweight binary classifier that identifies geospatial queries with $F_1=0.930$ (95\% bootstrap CI $0.909-0.947$) and can be applied to web search query corpora beyond MS MARCO. Third, we derive and release a taxonomy of 88 geospatial query categories, illustrated in Figure~\ref{fig:clusters}, that quantifies the relative prevalence of distinct query types and provides a concrete empirical characterisation of geospatial web search at scale.

We make the code available on GitHub\footnote{https://github.com/ilyankou/ms-marco-geospatial} under the MIT license for reproducibility.

%% file: 2-related-work.tex
\section{Related work}

\subsection{Taxonomies of geographic questions}

Hamzei et al. mined MS MARCO \cite{bajaj_ms_2018} to characterise place questions by place type and scale \cite{hamzei_initial_2019}, and later by a broader semantic schema covering place names, types, activities, spatial relationships, and qualities \cite{hamzei_place_2019}. Xu et al. showed that professional geographic questions drawn from GIScience literature are syntactically richer and semantically distinct from web queries, centring on analytical entities such as distribution, pattern, and density \cite{xu_extracting_2020}. Another top-down approach by Kuhn et al. derived question templates from a spatial ontology and a taxonomy of place facets, arguing that corpus-driven methods inevitably under-represent emotive and physical facets due to collection bias \cite{kuhn_semantics_2021}.

On the system side, Punjani et al. addressed answering geographic natural language questions over structured linked data, translating questions into \textsc{GeoSPARQL} queries via handcrafted templates. Their \textsc{GeoQuestions201} benchmark defines seven question categories of increasing structural complexity, covering topological, cardinal direction, and distance relations \cite{punjani_template-based_2018}. Kefalidis et al. extended this line of work with \textsc{GeoQuestions1089}, a larger benchmark, and showed that even improved template-based systems leave substantial headroom on geographic question answering \cite{kefalidis_question_2024}.

\subsection{Geospatial content in search queries and on the web}

Several studies have attempted to quantify how much of online information and search activity is geographic in nature, using methods that range from keyword matching to network analysis to LLM-based classification.

Sanderson and Kohler found that roughly 18.6\% of 1 million Excite\footnote{Major web portal and search engine of the late 1990s} search queries studied contained a geographic term, with spatial relationship terms appearing rarely, suggesting users at the time did not expect search engines to handle relational spatial queries \cite{sanderson_analyzing_2004}. Jones et al. estimated place-name prevalence of Yahoo queries at 12.7\%; they found that the vast majority target cities, and demonstrated that acceptable query-to-target distance is strongly topic-dependent, with restaurant queries clustering within tens of kilometres while accommodation queries tolerate far greater distance \cite{jones_geographic_2008}. Gan et al. analysed the AOL dataset of 36 million records from 2006\footnote{The dataset was withdrawn shortly after release following privacy concerns}; the authors built a geo/non-geo classifier and identified 13.39\% of geographic-intent queries, and suggested a taxonomy of 23 broad and high-level categories that `combine aspects of topicality and desired type of interaction', such as `Tourism/Travel', `Medical', `Entertainment', or `Navigational' \cite{gan_analysis_2008}. Henrich and Lüdecke also classified AOL user intent, and found habitation, accommodation, spare time, and information as dominant concepts; 65\% of queries implied physical travel to the target; and the majority were selective rather than covering, seeking a specific point rather than documents spanning an area \cite{henrich_characteristics_2007}.

Beyond queries, Hahmann and Burghardt tested the much-cited claim that 80\% of all information is geospatially referenced, finding via network and cognitive analyses of the German Wikipedia that the defensible figure is closer to 56--59\% \cite{hahmann_how_2013}. A more recent analysis of Common Crawl found that 18.7\% of web documents contain explicit geospatial information such as coordinates or addresses \cite{ilyankou_quantifying_2024}, remarkably close to Sanderson and Kohler's query-side estimate two decades earlier.

\subsection{Place theory and information needs}

Edwardes and Purves showed that spatial vocabularies are strongly locale-sensitive; British contributors preferred \emph{hill} to \emph{mountain}, and human settlements dominated over physical geography, implying that any fixed keyword list will misclassify a substantial portion of implicitly spatial queries \cite{edwardes_theoretical_2007}. Purves et al. argued more fundamentally that existing information systems represent place in an impoverished way, reducing it to named points or coordinate-bound objects, and that place is instead vague, relational, and socially constructed \cite{purves_places_2019}. Therefore, it would follow that a query about living costs is just as place-anchored as any other query containing a toponym. Shanon \cite{shanon_answers_1983} demonstrated that \emph{where-question} responses are governed not only by containment hierarchies and physical distance but by the cognitive salience of referents and shared epistemic context, dimensions that template-driven geographic information retrieval systems do not fully address to this day.

\subsection{Limitations of prior work}

Previous query log studies and taxonomy building attempts share a critical limitation: they characterise geospatial queries by their syntactic form or semantic encoding, not by what people are actually trying to find out. Even sophisticated approaches to detecting geospatial language in text, such as spatial role labelling of geospatial prepositions \cite{radke_detecting_2019}, rely on surface-form signals and thus miss implicitly geospatial queries. Most rely on pre-filtered subsets: explicitly location-labelled records, toponym-containing queries, or purpose-built corpora that by design exclude such queries, producing prevalence estimates that are systematically too low when filtering on toponyms or place names (12.7\% \cite{jones_geographic_2008}, 13.39\% \cite{gan_analysis_2008}), or that conflate surface form with intent when matching geographic terms more broadly. What remains unknown is the topical landscape of geospatial web search: not whether a query contains a place name or a spatial predicate, or how people phrase their queries, but whether people are searching for nearby restaurants, flight times between cities, the best places to move to, local weather, or regional costs of living. We address this gap by characterising geospatial queries by topical intent rather than syntactic form, applied without pre-filtering to a corpus of 1.01 million web search queries.

%% file: 3-methodology.tex
\section{Methodology}

\begin{figure}[t]
  \centering
  \includegraphics[width=\linewidth]{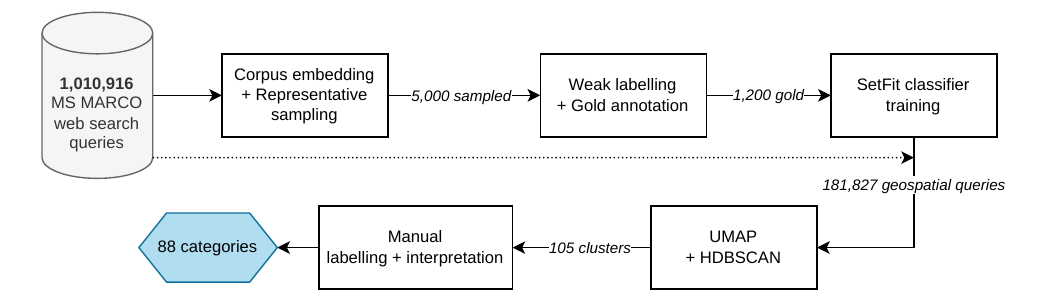}
  \caption{Overview of the methodology. Starting from the full MS MARCO corpus of 1.01M queries, we sample 5,000 representative queries for gold annotation, train a SetFit classifier to identify 181,827 geospatial queries, and apply UMAP dimensionality reduction and HDBSCAN clustering on geospatial query embeddings to derive a taxonomy of 88 geospatial query categories, grouped into 9 themes.}
  \label{fig:pipeline}
\end{figure}

Figure~\ref{fig:pipeline} summarises the full methodology, from corpus assembly through to taxonomy derivation.

\subsection{Defining \emph{geospatial}}

No consensus definition of a spatial, geospatial, or geographic question or query exists. In the field of geographical question answering, Mai et al. broadly define \emph{geographic} questions as those `involving geographic entities, geographic concepts, or spatial relations as parts of the natural language questions' \cite{mai_geographic_2021}, and Kefalidis et al. define \emph{geospatial} questions as those `requiring qualitative or quantitative geographic knowledge to be answered' \cite{kefalidis_question_2024}. We thus define a \textit{geospatial query} as follows:

\begin{quote}
    A query is geospatial if it requires qualitative or quantitative geographic knowledge of Earth-bound features to be answered. This is usually the case if the query involves a geographic entity, a geographic concept, or a spatial relation.
\end{quote}

In our definition, we exclude questions that are anatomical, microscopic, or astronomical in scale, and fictional or abstract `where' questions.

\subsection{Identifying geospatial queries}

We combined all available MS MARCO search queries\footnote{Available to download at https://msmarco.z22.web.core.windows.net/msmarcoranking/queries.tar.gz} from the train, validation, and test subsets to produce a single corpus of 1,010,916 unique web search queries, each represented as a short\footnote{Mean query length is 35 characters, and 99.9\% of queries are under 131 characters}, typically lowercased natural-language string, ranging from clearly not geospatial (e.g., `what is the human main muscles', \textit{query ID 825350}), to clearly geospatial (e.g., `what county is badger mn in?',  \textit{602750}), to many contestable, such as `most western point in portugal' (\textit{459663}), `how many square feet in an acre/' (\textit{296354}), and `what all places do i need to change my address when i move' (\textit{553148}). 

To support sampling and downstream clustering, we pre-compute 384-dimensional embeddings for all 1.01M search queries using a compact, general-purpose text embedding model \textit{BAAI/bge-small-en-v1.5}\footnote{Available on HuggingFace: https://huggingface.co/BAAI/bge-small-en-v1.5} via Sentence Transformers \cite{reimers_sentence-bert_2019}, which have been shown to encode some quasi-geospatial structure \cite{ilyankou_sentence_2024}. The chosen model offers strong semantic similarity performance for short English texts and is computationally feasible on a consumer laptop; the BGE family generally performs competitively on the Massive Text Embedding Benchmark (MTEB)\footnote{The leaderboard is available at https://huggingface.co/spaces/mteb/leaderboard} \cite{muennighoff_mteb_2023}.

\subsubsection{Constructing a gold dataset}

We sample 5,000 queries from the entire MS MARCO corpus for labelling; exploratory manual verification suggested this would yield at least 500 positive (i.e., geospatial) samples. We use \textit{k-means++}\footnote{https://scikit-learn.org/stable/modules/generated/sklearn.cluster.kmeans\_plusplus.html} \cite{arthur_k-means_2007} to generate 5,000 centroids in the embedding space and snap each to its nearest query. Compared to random sampling, we expect this to produce broader coverage of rare query types rather than over-representing dense clusters.

For each sampled query, we first obtain a \textit{weak} label using a locally hosted Llama-3.1 \cite{grattafiori_llama_2024_manual} via Ollama\footnote{https://ollama.com/library/llama3.1} using the few-shot prompt shown in the Appendix \ref{appendix:weak-label-prompt}. The model is computationally efficient to run locally, and its instruction-following capability is sufficient for binary classification of short natural-language strings. To reduce the impact of stochastic variation in LLM outputs, we run the model 5 times per query at a low sampling temperature of 0.3 and assign the majority vote as the weak label. The lead author then manually verifies and corrects the weak labels to produce the gold dataset of $1,200$ queries, which we release on GitHub\footnote{https://github.com/ilyankou/ms-marco-geospatial/tree/main/gold-dataset}. To assess the reliability of the lead author's gold labels, the remaining two authors independently of each other annotate a 200-query sample drawn from the gold set; we report pairwise Cohen's $\kappa$ \cite{cohen_coefficient_1960} for inter-annotator agreement across three annotator pairs.

\subsubsection{Training a SetFit binary classifier}

We use the gold dataset to contrastively train a lightweight geospatial/non-geospatial binary classifier using SetFit \cite{tunstall_efficient_2022}, a few-shot learning framework that fine-tunes a sentence embedding model via contrastive learning before fitting a classification head for strong performance with limited labelled data. We use \textit{BAAI/bge-small-en-v1.5} as the base sentence-transformer. SetFit is designed for few-shot learning and requires few labelled examples to perform well \cite{tunstall_efficient_2022}; we therefore split the dataset with stratification into train (200 samples), validation (200), and hold-out test (800) partitions for a robust initial evaluation. We train for five epochs with a batch size of 64 and a learning rate of $2\times10^{-5}$, keeping the SetFit default of $20$ contrastive pair-sampling iterations per epoch. We evaluate every epoch and apply early stopping after 2 rounds with no improvements in embedding loss. The model is evaluated on the hold-out test set; we report 95\% confidence intervals computed via 1,000 bootstrap resamples of the test set. For production inference (i.e., to label the 1.01M queries as geospatial or not), we retrain the same configuration on the full gold dataset for 3 epochs, corresponding to the best validation checkpoint observed during initial evaluation. We release the trained classifier on HuggingFace\footnote{https://huggingface.co/ilyankou/is-geospatial-query}.

\subsection{Building taxonomy}

Once all geospatial queries are identified, we use clustering on their embeddings and manual interpretation to categorise geospatial web search and identify key themes.

\subsubsection{Dimensionality reduction and density-based clustering}

We take the subset of queries predicted as geospatial and retrieve their corresponding $384$-d embeddings. Because density-based clustering in high dimensions is difficult (the problem often referred to as the \emph{curse of dimensionality} \cite{assent_clustering_2012}), we first reduce dimensionality using the Uniform Manifold Approximation and Projection algorithm, or UMAP\footnote{https://umap-learn.readthedocs.io/en/latest/} \cite{mcinnes_umap_2020}, which projects high-dimensional embeddings into a lower-dimensional space while preserving local neighbourhood structure, and then cluster the reduced representations using Hierarchical Density-Based Spatial Clustering of Applications with Noise, or HDBSCAN\footnote{https://scikit-learn.org/stable/modules/generated/sklearn.cluster.HDBSCAN.html} \cite{campello_density-based_2013}. This combination is widely used for topic modelling over transformer embeddings \cite{angelov_top2vec_2020, sia_tired_2020, grootendorst_bertopic_2022, allaoui_considerably_2020}.

\subsubsection{Grid search over UMAP and HDBSCAN parameters}

Density-based clustering relies on density thresholds whose appropriate values depend on the underlying data distribution; selecting a suitable density level is inherently difficult and data-dependent \cite{campello_density-based_2020}. We therefore perform a grid search over an interpretable parameter space for both UMAP and HDBSCAN to identify a set of parameters that produces a reasonable number of high-quality clusters. For UMAP, we vary (i) output dimensionality $\in \{5, 10, 15\}$ and (ii) neighbourhood size $\in \{10, 25, 50\}$. For HDBSCAN, we vary (iii) minimum cluster size $\in \{25, 50, 100, 200\}$ and (iv) minimum number of samples, set to $\{0.2, 0.5, 1.0\} \times $ \textit{minimum cluster size}, using the default Excess of Mass (\textit{`eom'}) cluster selection method. This gives us $3 \times 3 \times 4 \times 3 = 108$ configurations, each evaluated with a fixed random seed (42).

For each configuration, we compute four quality metrics: (a) the Density-Based Clustering Validation (DBCV) score \cite{moulavi_density-based_2014}, a density-aware measure of within-cluster cohesion and between-cluster separation suited to the arbitrary-shape clusters produced by methods such as HDBSCAN (range $-1$ to $1$, higher is better), (b) the noise fraction, i.e. the proportion of queries assigned to cluster `$-1$'), (c) the number of non-noise clusters, and (d) the median non-noise cluster size. We use DBCV rather than silhouette scores, as the latter assumes convex clusters and has no principled treatment of noise points, both of which make it unsuitable for evaluating HDBSCAN output.

\subsubsection{Consistency checks}

UMAP is stochastic, and its randomness can affect downstream clustering. To assess how stable our clustering is, we pick the top-10 configurations by DBCV that produced at least 10 clusters (to exclude degenerate solutions with too few clusters to be informative), and re-run each across six random seeds $\in \{0, 1, 2, 3, 4, 42\}$. For each configuration, we report the mean and standard deviation of DBCV, noise fraction, and cluster count across seeds, and select the configuration with the strongest combination of high mean DBCV and low cross-seed variance. For final taxonomy extraction, we run the full pipeline once on all 181,827 queries using the selected configuration.

\subsubsection{Cluster interpretation}

To support manual naming of resulting clusters, for each cluster we (i) identify a representative query, defined as the query whose embedding is closest to the cluster centroid, (ii) extract salient unigram and bigram terms using Term Frequency-Inverse Document Frequency (TF-IDF) over concatenated cluster texts, with English stop-words removed, and (iii) randomly sample 10 queries per cluster to provide additional context for annotators. These summaries are exported as Markdown and as a spreadsheet for manual annotation by the authors.

Following manual annotation, we merge clusters that we judge to represent the same query intent (most notably, multiple US state-specific `what county is [place] in' clusters fragmented by named entities rather than intent), yielding a final taxonomy of 88 categories.

We group the categories into nine broad themes -- Statistical, Temporal, `POIs and Commercial', Administrative, Physical, Cultural, Historical, Biographical, and Events -- for organisational convenience rather than as a theoretically grounded hierarchy. We experimented with alternative groupings but found that meaningful structure resides in the leaf categories themselves; any higher-level grouping risks implying a cleaner taxonomy than the data supports.

The resulting taxonomy is exported as a parent--child JSON graph for visualisation.

%% file: 4-results.tex
\section{Results}

\subsection{Gold dataset}

Of the 5,000 sampled queries, 628 received at least one `geospatial' vote from the LLM, of which the vast majority (547) were unanimously labelled `geospatial' across all five LLM runs. We manually verified 561 true weak positives and identified 67 false weak positives. We also randomly sampled and manually verified 572 weak negatives, identifying 7 false weak negatives. Thus, the gold dataset consists of 568 positives and 632 negatives (total size of $1,200$). 

\begin{figure}[h]
\includegraphics[width=1.0\textwidth]{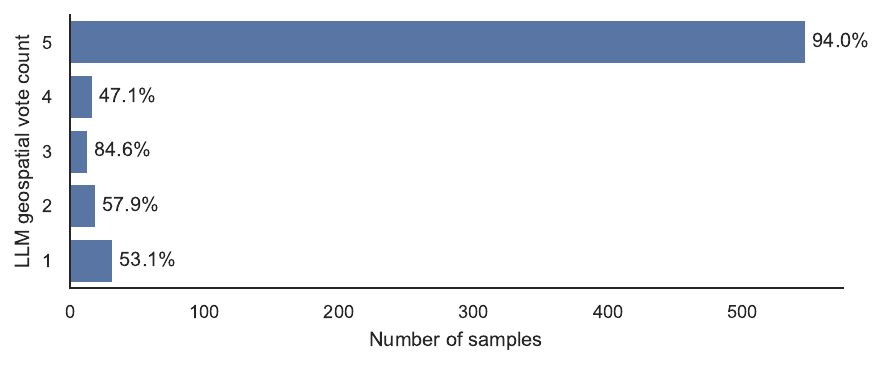}
\caption{Human verification of weak geospatial labels across LLM vote counts. Bars indicate the number of queries receiving 1-5 spatial votes across repeated LLM runs; percentages denote the proportion confirmed as geospatial after manual verification.}
\label{fig:spatial-1-5}
\end{figure}

Figure~\ref{fig:spatial-1-5} shows human agreement with the weak LLM labels as a function of the number of `geospatial' votes (between 1 and 5). The distribution is heavily skewed toward unanimous predictions, suggesting that most weakly-positive queries are unambiguous for the Llama-3.1 model under the prompt and the definition adopted. The intermediate bins (2-4 votes) contain few queries, so their confirmation rates should not be read as a monotonic trend. Unsurprisingly, the more obvious geospatial queries, such as `what county is springfield, or' (\textit{613049}) or `which region is egypt located' (\textit{1018537}) typically received all 5 positive votes; more borderline cases, such as `biggest house in the world price' (\textit{53174}) or `where do people think dolphins live' (\textit{971381}) produced inconsistent votes across runs, with the model split between geospatial and non-geospatial judgements.

To assess inter-annotator agreement, the second and third authors independently re-annotated 200 samples drawn from the gold set. The lead, second and third authors all agreed on 179 of 200 samples (89.5\%); of those, 107 were negative and 72 positive. Pairwise Cohen's $\kappa$ was 0.88 (lead vs second author), 0.84 (lead vs third), and 0.84 (second vs third), indicating `almost perfect' agreement \cite{landis_measurement_1977}. As expected, disagreements concentrated on borderline cases. For example, only the lead author read `va' as Virginia in the query `va tax contact' (\textit{535301}); `cricket wireless address' (\textit{112731}) was labelled as geospatial by the lead and second authors but not the third; `biggest house in the world price' (\textit{53174}) was labelled as geospatial by the lead and third authors, but not the second.

\subsection{SetFit classifier}

When trained on just 200 samples (105 negative and 95 positive) and tested on 800 samples (421 negative and 379 positive) of the gold dataset, the SetFit binary classifier achieves an accuracy of 0.934 (95\% bootstrap CI $0.915-0.950$) and an $F_1$ score of 0.930 (95\% bootstrap CI $0.909-0.947$), producing 27 false negatives and 26 false positives. The majority of misclassified samples are borderline geospatial and can be argued both ways; for example, `biggest snakes in world' (FP, \textit{53491}), `when is the good time to see northern lights' (FP, \textit{952893}), `address for expresspcb' (FN, \textit{11492}), or `how wide is the base of the great wall' (FN, \textit{1175805}). This result is particularly notable given the small size of the training set and near-equal class distribution in both training and test sets, making $F_1$ a meaningful performance indicator rather than a result of class imbalance.

We further verify the classifier's robustness by sense-checking its behaviour on made-up, semantically complex queries and short phrases, a subset of which is demonstrated in Table~\ref{tab:sense-check}.

\input{tables/sense-check}

After retraining the classifier on the full gold dataset consisting of $632$ negative and $568$ positive cases, we run it on the entire MS MARCO dataset consisting of $1,010,916$ queries, and identify $181,827$ spatial queries, or $18.0\%$ of the entire dataset.

An analysis of the most frequent first words in each class (see Table~\ref{tab:first-words}) reveals that \textit{what} dominates both geospatial and non-geospatial queries (29.6\% and 36.1\% respectively), \emph{how} appears in the top-3 for both (11.6\% and 17.9\%), while \textit{where} ranks second in geospatial queries (15.8\%) but only 14th in non-geospatial queries (0.8\%). The overlap between the two vocabularies is striking: 14 of the top-20 first words appear in both lists.

\input{tables/first-words.tex}

\subsection{Grid search for clustering}

The top-10 UMAP and HDBSCAN configurations by DBCV, which produced at least 10 clusters, are shown in Table~\ref{tab:top10config}.

\input{tables/top10config.tex}

Table~\ref{tab:consistency} reports the consistency of the top-10 configurations across six repeated runs with different random seeds. Six configurations are stable, with standard deviation for both DBCV and noise fraction below $0.050$.

\input{tables/consistency.tex}

Based on the consistency analysis, configurations \#21 and \#23 are superior; we pick configuration \#21 as it achieves a lower mean share of noise ($0.389$ vs $0.409$ in configuration \#23) and a higher number of clusters (mean $108$ vs $75$) despite having a slightly lower mean DBCV ($0.310$ vs $0.326$). For the seed, we pick 42, whose DBCV falls near the configuration's median across runs (we deliberately avoid selecting the highest-DBCV seed to prevent cherry-picking).

\subsection{Final clustering}

\begin{figure}[ht]
    \centering
    \includegraphics[width=1.02\linewidth]{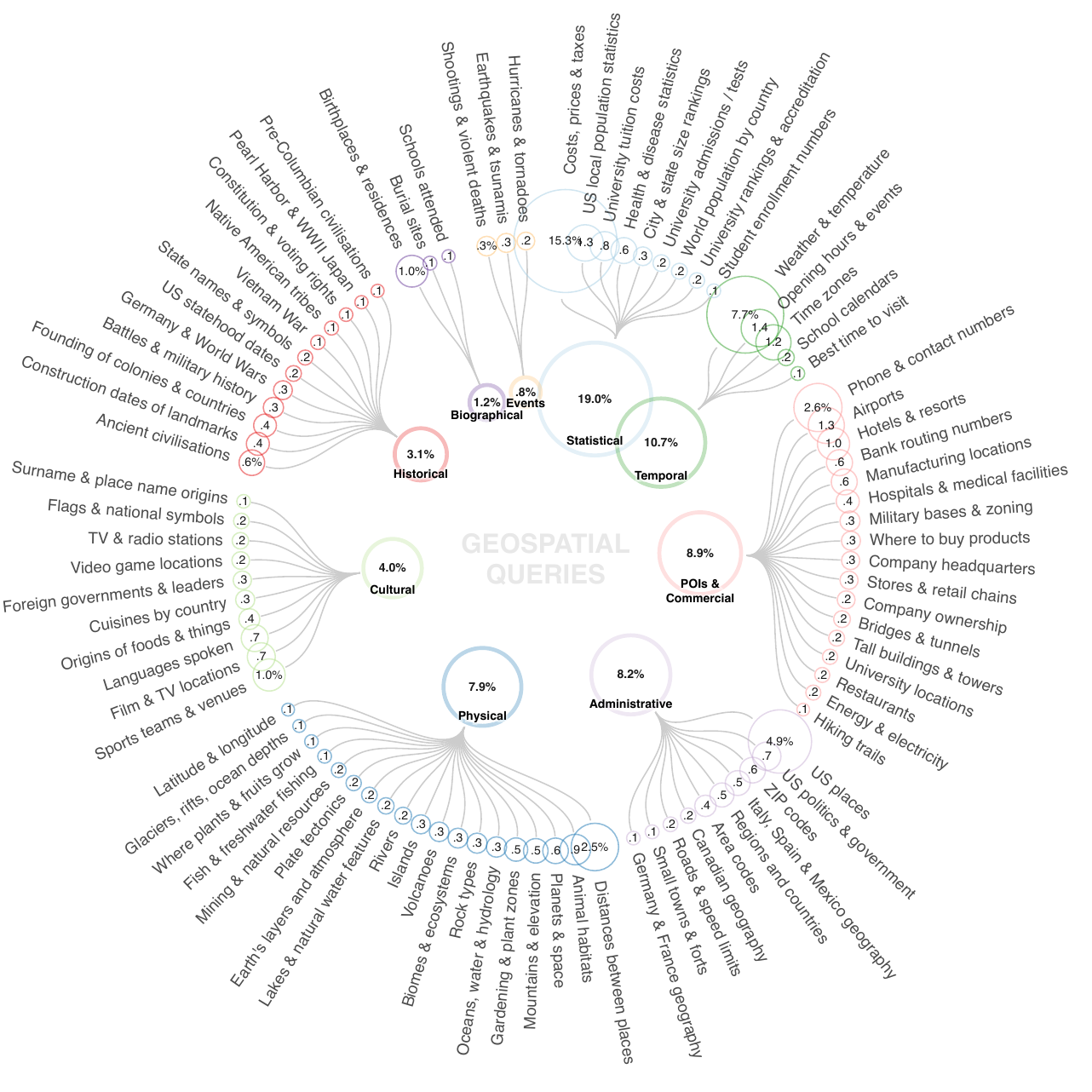}
    \caption{Geospatial query clusters grouped into nine themes. Unclustered (noise) queries representing 36.2\% of identified geospatial queries are not shown.}
    \label{fig:clusters}
\end{figure}

The radial hierarchy chart in Figure~\ref{fig:clusters} illustrates 88 categories (collapsed from 105 initial clusters), grouped into nine broad themes.

Examples of clusters' representative queries, top-20 most frequently occurring terms, and random samples of 10 queries are illustrated in Table~\ref{tab:clustering-info-sample}. Table~\ref{tab:cluster-rep-queries} in the Appendix shows resulting labelled clusters with sizes and representative queries. We make all cluster-level data available on GitHub\footnote{https://github.com/ilyankou/ms-marco-geospatial/blob/main/interim/cluster\_review.md}.

\input{tables/clustering-info-sample}

%% file: tables/sense-check.tex
\begin{table}[ht]
\centering
\begin{tabular}{ll}
\toprule
\textbf{Correctly classified as geospatial} & \textbf{Correctly classified as non-geospatial} \\
\midrule
nearest hospital & near impossible \\
countries bordering ukraine  & borderline invisible \\
distance from paris to berlin & keep your distance \\
restaurants along the m1 & stop it m8 \\
where should i live & where can i escape mentally \\
flood risk in this area & not my area of expertise \\
is it a long flight london to kaunas & a long way to go in my career \\
\bottomrule
\end{tabular}
\caption{Some examples of correctly classified, semantically complex made-up queries and phrases (\emph{not} part of MS MARCO) to sense-check the trained classifier's behaviour.}
\label{tab:sense-check}
\end{table}

%% file: tables/first-words.tex
\begin{table}[ht]
\centering
\small
\begin{tabular}{clr clr | clr clr}
\toprule
\multicolumn{6}{c}{\textbf{Geospatial}} & \multicolumn{6}{c}{\textbf{Non-geospatial}} \\

\midrule

\textbf{\#} & \textbf{Word} & \textbf{\%} & \textbf{\#} & \textbf{Word} & \textbf{\%} & \textbf{\#} & \textbf{Word} & \textbf{\%} & \textbf{\#} & \textbf{Word} & \textbf{\%} \\
\midrule
1 & what    & 29.6 & 11 & cost & 1.0        & 1 & what & 36.1 & 11 & define & 1.0 \\
2 & where   & 15.8 & 12 & population & 0.9  & 2 & how & 17.9 & 12 & definition & 1.0 \\
3 & how     & 11.6 & 13 & what's & 0.9      & 3 & who  &  3.7 & 13 & cost & 0.9 \\
4 & average &  3.4 & 14 & does & 0.6        & 4 & is   &  3.0 & 14 & where & 0.8 \\
5 & when    &  3.3 & 15 & can & 0.5         & 5 & when &  2.6 & 15 & do & 0.7 \\
6 & weather &  2.8 & 16 & most & 0.5        & 6 & can &  2.1 & 16 & the & 0.6 \\
7 & is      &  2.5 & 17 & largest & 0.5     & 7 & why &  1.8 & 17 & are & 0.6 \\
8 & which   &  1.9 & 18 & distance & 0.5    & 8 & which &  1.8 & 18 & meaning & 0.5 \\
9 & who     &  1.8 & 19 & temperature & 0.5 & 9 & does &  1.3 & 19 & what's & 0.4 \\
10 & why    &  1.1 & 20 & the & 0.4         & 10 & average & 1.2 & 20 & causes & 0.4 \\
\bottomrule
\end{tabular}
\caption{Top-20 most frequent first words in geospatial and non-geospatial MS MARCO queries.}
\label{tab:first-words}
\end{table}

%% file: tables/top10config.tex
\begin{table}[ht]
\centering
\footnotesize
\begin{tabular}{c|rr|rr|rrrr}
\toprule
& \multicolumn{2}{c|}{\textbf{UMAP}} & \multicolumn{2}{c|}{\textbf{HDBSCAN}} & & & & \\
\textbf{\makecell{Config\\ID}} & \textbf{Dims} & \textbf{Neigh} & \textbf{\makecell{Min \\cluster\\size}} & \textbf{\makecell{Min\\samples}} & \textbf{DBCV} & \textbf{\makecell{Noise}} & \textbf{\makecell{Num\\clusters}} & \textbf{\makecell{Median\\cluster\\size}} \\
\midrule

93 & 15 & 25 &	200 &	40 &	0.358 & 0.359 	& 102 & 449.5 \\
79 & 15 & 10 &	100 &	50 &	0.341 & 0.359 &	182 & 246 \\
81 & 15 & 10 &	200 &	40 &	0.337 & 0.388 &	122 & 476 \\
22 & 5  & 25 &	200 &	100 &	0.334 & 0.389 &	91 	& 467 \\
21 & 5  & 25 &	200 &	40 	& 0.333 & 0.362 &	105 & 469 \\
95 & 15 & 25 &	200 &	200 & 0.324 & 0.417 &	77 	& 553 \\
9  & 5  & 10 &	200 &	40 & 0.319 	& 0.392 &	122 & 443.5 \\
45 & 10 & 10 &	200 &	40 & 0.317 	& 0.374 &	114 & 475 \\
23 & 5  & 25 &  200 &	200 & 0.315 & 0.426 &	79 &  485 \\
76 & 15 & 10 & 	50 	&    25 & 0.312 & 0.418 &	415 &  119 \\
 
\bottomrule
\end{tabular}
\caption{Top-10 UMAP and HDBSCAN hyper-parameter configurations by DBCV score, restricted to those producing at least 10 clusters. \textit{Dims} is UMAP output dimensionality, \textit{Neigh} is UMAP \texttt{n\_neighbors}, \textit{Min cluster size} and \textit{Min samples} are HDBSCAN's \texttt{min\_cluster\_size} and \texttt{min\_samples} parameters; \textit{Noise} is the proportion of queries assigned to the noise (`$-1$') cluster.}
\label{tab:top10config}
\end{table}

%% file: tables/consistency.tex
\begin{table}[ht]
\centering
\footnotesize
\begin{tabular}{c|cccc|cccc|rrrr}
\toprule
& \multicolumn{4}{c|}{\textbf{DBCV}} & \multicolumn{4}{c|}{\textbf{Noise}} & \multicolumn{4}{c}{\textbf{Num clusters}} \\
\textbf{\makecell{Config\\ID}} & \textbf{Min} & \textbf{Mean} & \textbf{Max} & \textbf{Std} & \textbf{Min} & \textbf{Mean} & \textbf{Max} & \textbf{Std} & \textbf{Min} & \textbf{Mean} & \textbf{Max} & \textbf{Std} \\
\midrule
93 & 0.245 & 0.364 & 0.712 &	0.176 &	0.000 &	0.332  & 0.432 & 0.165 & 3 & 88 & 109 	& 41.9 \\
45 & 0.097 & 0.363 & 0.811 &	0.242 &	0.000 &	0.245  & 0.400 & 0.190 & 4 & 77 & 115 &	56.3\\
23 & 0.279 & 0.326 & 0.416 &	0.049 &	0.341 &	0.409  & 0.464 & 0.040 & 64 & 75 & 82 &	6.4\\
21 & 0.257 & 0.310 & 0.367 &	0.040 &	0.358 &	0.389  & 0.415 & 0.027 & 100 &	108 & 114 &	6.2\\
81 & 0.178 & 0.306 & 0.358 &	0.069 &	0.001 &	0.244  & 0.388 & 0.189 & 4 &	77 	& 122 &	56.6\\
95 & 0.227 & 0.289 & 0.348 &	0.044 &	0.410 &	0.432  & 0.470 & 0.026 & 72 &	76 	& 82 &	3.4\\
79 & 0.247 & 0.282 & 0.341 &	0.033 &	0.344 &	0.395  & 0.442 & 0.038 & 173 &	195 & 208 	&14.3\\
76 & 0.238 & 0.279 & 0.312 &	0.031 &	0.412 &	0.421  & 0.443 & 0.011 & 406 &	421 & 459 	&19.4\\
9  & 0.092 & 0.271 & 0.412 &	0.109 &	0.000 &	0.253  & 0.406 & 0.196 & 4 &	79 & 122 	&58.3\\
22 & 0.231 & 0.269 & 0.334 &	0.040 & 0.389 & 0.434  & 0.461 & 0.033 & 87 	& 96 & 101 	&5.4 \\

\bottomrule
\end{tabular}
\caption{Consistency of top-10 configurations across six repeated runs with various random seeds $\{0,1,2,3,4,42\}$, reporting minimum, mean, maximum, and standard deviation of the DBCV scores (1=perfect), noise fraction, and number of produced clusters.}
\label{tab:consistency}
\end{table}

%% file: tables/clustering-info-sample.tex
\begin{table}[ht]
\centering
\footnotesize
\begin{tabular}{p{2.0cm}  p{5.5cm}  p{5.5cm}}
\toprule

 & \textbf{Cluster \#48} & \textbf{Cluster \#55} \\
\midrule 

No. samples & 344 & 1015 \\

\midrule

Representative query
    & what is the tallest building in the world
    & what is the county of florida
    \\

\midrule 
    
Top-20 terms
    & tower, building, pisa, tower pisa, tallest, leaning, leaning tower, tall, tallest building, empire state, floors, building world, empire, state building, world, towers, khalifa, burj, burj khalifa, skyscraper
    
    & fl county, fl, county, florida, florida county, beach, palm, beach fl, population, florida located, jacksonville, fl population, county florida, beach florida, located, county palm, springs, orange, county jacksonville, naples
    \\

\midrule

Ten sample queries

    & largest moving structure in the world
    $\cdot$ what is the pyramid shaped building on the skyline of san francisco
$\cdot$ leaning tower of pisa facts
$\cdot$ how many floors is tower one
$\cdot$ how many floors does the tallest building in sydney have
$\cdot$ leaning tower locale
$\cdot$ how many feet is the world's tallest building
$\cdot$ how tall is victory tower
$\cdot$ how many floors in the empire state building.
$\cdot$ where is cornerstone condos located
    
    & homes for sale st. cloud fl.
$\cdot$ where is dunedin florida
$\cdot$ where is naples manor fl
$\cdot$ what county is atlantic beach fl in
$\cdot$ what county is harbor springs in
$\cdot$ where is legoland florida located
$\cdot$ what county is ft. lauderdale
$\cdot$ homes for sale jacksonville nc
$\cdot$ where is citrus florida located
$\cdot$ what county is holt fl
    \\

\midrule

Cluster label
    & Tall buildings \& towers
    & County membership and places in Florida $\rightarrow$ US places \\

\midrule 
Theme   & POIs \& Commercial
        & Administrative \\
\bottomrule 

\end{tabular}
\caption{To derive initial cluster labels, we looked at each cluster's representative query, top-20 most frequent terms, and ten randomly sampled queries.}
\label{tab:clustering-info-sample}
\end{table}

%% file: 5-discussion.tex
\section{Discussion}

\subsection{Characteristics of the taxonomy}

This study identifies 181,827 geospatial queries in MS MARCO, representing 18.0\% of the dataset, which is a substantially higher proportion than the 6.17\% of queries labelled as \emph{Location} in the original MS MARCO annotations \cite{bajaj_ms_2018}. This discrepancy reflects the breadth of our geospatial definition, which captures implicitly location-anchored queries such as weather conditions, prices, and biomes -- categories that may not be caught by keyword-based approaches.

18 of 105 initial clusters (\textit{\#55}, and \textit{\#88--\#104} in Table~\ref{tab:cluster-rep-queries} in the Appendix) largely represent variants of the query `what county is [place] in', differentiated primarily by US state abbreviation. This fragmentation is partly an HDBSCAN artefact as embeddings likely treat state abbreviations, such as `tx' or `ny', as distinguishing features. We group these fragmented clusters into a single \emph{US places} category for Figure~\ref{fig:clusters}. HDBSCAN treats $36.2\%$ of identified geospatial queries as noise (not shown in Figure~\ref{fig:clusters}), a substantial but unsurprising proportion given the scale and messiness of the real-world dataset. Natural-language web search queries are inherently heterogeneous, and some degree of noise is expected and acceptable in density-based clustering applied at this scale. The \emph{Planets \& space} cluster (1,045 largely astronomical queries, which our definition excludes) likely reflects the binary classifier not having been exposed to such queries during training; representing only 0.6\% of all geospatial queries, its effect on our analysis is negligible.

We use nine broad themes -- Statistical, Temporal, `POIs and Commercial', Administrative, Physical, Cultural, Historical, Biographical, and Events -- for organisational convenience rather than as a theoretically grounded hierarchy. We experimented with alternative groupings but found that meaningful structure resides in the leaf categories themselves; any higher-level grouping risks implying a cleaner taxonomy than the data supports. Because we report theme shares over the full set of geospatial queries, including noise, the percentages are conservative; the relative ranking of themes holds only insofar as noise is not concentrated in particular intents, which we do not test here.

The taxonomy reveals that geospatial web search is dominated by transactional and practical lookups rather than by either administrative or physical geography. The Statistical theme alone accounts for 19.0\% of geospatial queries and the Temporal theme a further 10.7\%. The \emph{Costs, prices \& taxes} (15.3\%) cluster, with queries such as `what is the average household income of lake oswego' (\textit{1152138}) or `what is currency in brazil' (\textit{736719}), is comparable in size to the Physical (7.9\%) and Administrative (8.2\%) themes combined. Users overwhelmingly turn to web search for place-anchored facts about money, time, and weather; questions about landforms, hydrology, or jurisdictional boundaries -- traditional GIS territory -- form a comparatively small share of expressed information needs.

\subsection{Implications for system design}

A large share of the taxonomy consists of clusters whose answers are open, stable, and objective. For example, \emph{Distances between places} (e.g., `how far is it from hartford to new london', \textit{231318}) and \emph{ZIP codes} (e.g., `what is burlington,ky zip code', \textit{726469}) are answerable deterministically using geoparsing, spatial databases and a traditional GIS toolkit, including distance calculations and containment operations. Query categories such as \emph{Regions and countries} (e.g., `what is the continent is algeria located', \textit{1151645}), \emph{Bridges \& tunnels} (e.g., `where does golden gate bridge connect to', \textit{973130}) or \emph{Volcanoes} (e.g., `what type of volcano is mt.santorini', \textit{915164}) are largely answerable from structured knowledge contained in geographical knowledge graphs. Such types of queries do not require generative or probabilistic systems to be answered correctly. This suggests that a hybrid retrieval architecture of routing objectively answerable geospatial queries to structured geodata sources while reserving generative models for more complex or subjective queries could improve both accuracy and reliability compared to treating all geospatial queries uniformly.

Categories such as \emph{Best time to visit} (e.g., `most affordable times to visit hawaii', \textit{456340}), \emph{University rankings \& accreditation}  (e.g., `is codarts a good school?', \textit{406730}), and \emph{Origins of foods \& things} (e.g., `where does spaghetti bolognese come from', \textit{973905}) are more subjective as their answers depend on personal preferences and the source of information used to answer them. These are the clusters where Anglophone-Western geographic and cultural biases in both the MS MARCO dataset and any downstream system are most likely to surface. Systems trained to answer such queries risk encoding and amplifying the biases of their training data under the guise of objective geospatial information. This risk is sharpest in conversational interfaces, which can deliver an evaluative or unverified answer with the same fluency as a deterministic lookup \cite{ilyankou_scenic_2026}.

Temporal or POI-specific categories such as \emph{Opening hours \& events} (e.g., `what day does the garrett pool open', \textit{616654}), \emph{Phone \& contact numbers} (e.g., `northampton county tax department pa phone number', \textit{1092161}), or \emph{Restaurants} (e.g., `what restaurant serves both mexican and steak', \textit{1109148}) demand answers that are not only geographically specific but temporally volatile: opening hours, phone numbers, and menus change, and businesses close and relocate much more frequently than coastlines redraw themselves. Unlike physical geography, which changes slowly, institutional information goes out-of-date quickly, making it particularly challenging for static retrieval systems to serve reliably without access to up-to-date data. The stakes are uneven across clusters: an outdated restaurant menu or opening time may cause minor inconvenience, whereas obsolete information about hospital locations or emergency services could have serious consequences. The \emph{Weather \& temperature} category (e.g., `what is the weather in nashville tomorrow', \textit{853623}), and the Events theme, accounting for 0.8\% of identified geospatial queries and covering \emph{Hurricanes \& tornadoes} (e.g., `where there storms last night', \textit{1001188}), \emph{Earthquakes \& tsunamis} (e.g., `was there a earthquake in alaska', \textit{541104}), and \emph{Shootings \& violent deaths} (e.g., `who is the woman that beat up woman at stop light in houston texas', \textit{1042188}), highlight the relationship between spatial and temporal information needs, which is another argument for hybrid architectures that can handle APIs, news and live data feeds in addition to static geographic knowledge.

The most significant implication concerns emerging natural-language interfaces, including conversational chatbot search. If nearly one in five web search queries is geospatial, these systems require robust geospatial reasoning for a substantial fraction of real user traffic. Existing geographic benchmarks for LLMs have largely focused on factual and location reasoning \cite{bhandari_are_2023,manvi_geollm_2024,roberts_gpt4geo_2023,mooney_towards_2023}, with comparatively less attention to the transactional, temporal, and subjective queries that dominate our taxonomy. Costs and prices, opening hours, and travel recommendations are examples of query categories where retrieval-augmented and real-time architectures matter most, yet they are largely absent from current evaluation benchmarks. Our taxonomy offers a more representative target for both system design and benchmarking, grounded in what users actually ask rather than what is easiest to test.

\subsection{Limitations and future work}

MS MARCO was collected prior to 2018 and reflects search behaviour on Bing at that time; the distribution of geospatial query types has likely shifted as search interfaces, user habits, and geospatial services have evolved. The dataset over-represents Anglophone North American users, and this is reflected directly in the taxonomy, where \emph{US places} (4.9\%) accounts for vastly more place-specific queries than \emph{Italy, Spain \& Mexico geography}, \emph{Canadian geography}, and \emph{Germany and France geography} clusters combined (0.8\%). Future work should seek to validate and extend the taxonomy using more geographically diverse query corpora.

The original MS MARCO dataset specifically excludes queries `with navigational and other intents' \cite{bajaj_ms_2018}, meaning that a distinct and practically significant category of geospatial information need is not represented in our taxonomy. Our 18.0\% is therefore best read as a lower bound, since routing and directions (likely to be a significant geospatial category, even in web search) are excluded by construction.

The corpus reflects web search habits and as such consists of short and typically single-intent queries, which constrains the taxonomy. Longer, multi-intent interactions typical of conversational AI systems, voice assistants, or chatbots fall outside what this analysis captures. In such systems, geospatial intent may be implicit, distributed across dialogue turns, or conditioned on prior context in ways that single-query classification cannot detect. Whether the underlying information needs remain stable across interfaces, or shift as users adapt to conversational systems, is an open empirical question that we aim to address in future work.

Several directions emerge from this work. First, the taxonomy clusters could be linked to existing geospatial datasets and spatial operations (for example, mapping distance queries to routing APIs, or county membership queries to administrative boundary datasets), providing a concrete bridge between user intent and existing GIS infrastructure. Second, the binary geospatial classifier could be extended into a multi-class router that directs queries to category-specific retrieval pipelines. The unclustered `noise' queries, representing over a third of all identified geospatial queries, also deserve further study; repeating the full clustering pipeline on this subset alone may reveal latent structure that the current parameter settings were too coarse to resolve. Finally, the distinction between objective and subjective geospatial queries merits a more formal investigation, as it has direct implications for system design, bias mitigation, and the appropriate use of generative versus deterministic systems.

%% file: 6-conclusion.tex
\section{Conclusion}

Geospatial web search queries are acts of spatial cognition. When someone types `in which county is north hollywood california' (\textit{394665}) or `cost of living in turkmenistan' (\textit{105311}), they are navigating a conceptually structured space of places, attributes, and relationships. Our finding that $18.0\%$ of MS MARCO queries are geospatial under a semantically broad definition is evidence that geospatial thinking pervades everyday information behaviour at a scale that toponym-based studies in particular have underestimated.

The derived taxonomy shows that this everyday spatial cognition diverges from what GIS has traditionally modelled. Questions about landforms, hydrology, and physical geography are outnumbered by transactional and practical lookups about costs, weather, and opening hours. This has direct consequences for systems designed to serve such queries.

The taxonomy also surfaces a more fundamental split: queries about distances, ZIP codes, or county membership presuppose that place has determinate answers; queries about the best place to live or the right season to visit presuppose that place is evaluative and culturally inflected. These two modes -- geospatial information as fact versus geospatial information as judgement -- demand different infrastructure and raise different risks. Distinguishing them is, we argue, a precondition for handling geographic knowledge responsibly.

%% file: 7-appendix.tex
\clearpage

\appendix

\section{LLM prompt for weak labelling}\label{appendix:weak-label-prompt}

\lstdefinestyle{prompt}{
    basicstyle=\small\ttfamily,
    breaklines=true,
    frame=single,
    rulecolor=\color{gray!40},
    backgroundcolor=\color{gray!5},
    showstringspaces=false,
    columns=fullflexible,
    keepspaces=true,
}

\begin{figure}[ht]
\centering
\begin{lstlisting}[style=prompt, caption={Prompt for Llama-3.1 to label 5,000 sampled queries as geospatial/non-geospatial}, label={lst:weak-labelling-prompt}]
Classify the search query as geospatial (true) or not (false).

A query is geospatial if it requires qualitative or quantitative geographic
knowledge of Earth-bound features to be answered. This is usually the case if the query involves:
- A geographic entity (named place on Earth: city, country, river, POI, address)
- A geographic concept (place type: city, lake, mountain, park, building)
- A spatial relation (near, within, north of, between, borders, crosses, distance)

Non-geospatial: anatomical, microscopic, astronomical, fictional, or abstract
`where' questions; queries needing no geographic knowledge.

Output only: true or false.

Examples:
  * How far is Brighton from London -> true
  * what does square from greece mean -> false
  * Capital of France -> true
  * What does `Dutch courage' mean -> false
  * Population of Piraeus -> true
  * Where is the epimysium found -> false
  * Restaurants near Hyde Park -> true
  * Where did the name Missouri come from -> false
  * Where does Paris Hilton live -> true
  * What is a river -> false
  * What city is Ebright Azimuth in -> true
  * is the water underneath a hurricane calm -> false
  * how many state are in the us -> true
  * Where is the SSID -> false
  * where are new franchises needed -> true
  * which side is the us flag put -> false
  * what languages are spoken in israel -> true
  * Where is Hogwarts -> false
  * most western point in portugal -> true
  * What is Greek yoghurt -> false

Query: <SEARCH QUERY>
Answer:

\end{lstlisting}
\end{figure}

\newpage

\section{Clusters with representative queries}

\input{tables/cluster-rep-queries}

%% file: tables/cluster-rep-queries.tex
\begin{table}[!ht]
\centering
\footnotesize
\begin{tabular}{r r p{5cm} p{7.5cm}}
\toprule
\textbf{\#} & \textbf{Size} & \textbf{Label} & \textbf{Representative query \textit{(query ID)}} \\
\midrule

-1  &  65823  &  Unclustered (HDBSCAN noise)  &  what is the url for the state bank of the lakes \textit{(852466)}  \\
0  &  696  &  Area codes  &  area code numbers united states \textit{(26408)}  \\
1  &  1054  &  ZIP codes  &  what's the zip code \textit{(933999)}  \\
2  &  13966  &  Weather \& temperature  &  what's the temperature outdoors \textit{(933263)}  \\
3  &  27761  &  Costs, prices \& taxes  &  what is the current minimum wage for ny \textit{(1151495)}  \\
4  &  526  &  Earthquakes \& tsunamis  &  what is been the biggest earthquake \textit{(722978)}  \\
5  &  1344  &  US politics \& government  &  how many representatives for each state \textit{(294437)}  \\
6  &  596  &  Military bases \& zoning  &  what military base is the largest in the us \textit{(878154)}  \\
7  &  298  &  State names \& symbols  &  when was the state nickname chosen \textit{(962316)}  \\
8  &  1296  &  Film \& TV locations  &  where was where the heart is filmed \textit{(1004279)}  \\
9  &  217  &  Schools attended  &  what college did trump attend \textit{(597060)}  \\
10  &  1744  &  Birthplaces \& residences  &  where was mlk born \textit{(1002963)}  \\
11  &  505  &  City \& state size rankings  &  what is the largest city in the US by area \textit{(827081)}  \\
12  &  818  &  Hospitals \& medical facilities  &  university of iowa hospitals and clinics patient \textit{(532623)}  \\
13  &  316  &  Canadian geography  &  toronto is in what province \textit{(522784)}  \\
14  &  441  &  Hurricanes \& tornadoes  &  where does a hurricane occur \textit{(972347)}  \\
15  &  1406  &  University tuition costs  &  usu tuition cost per semester \textit{(534985)}  \\
16  &  1096  &  Bank routing numbers  &  us bank routing numbers \textit{(533522)}  \\
17  &  4641  &  Phone \& contact numbers  &  state street service desk number \textit{(503076)}  \\
18  &  246  &  Best time to visit  &  when is the best time to visit the bahamas \textit{(952523)}  \\
19  &  287  &  Roads \& speed limits  &  types of roads and their speed limits \textit{(529594)}  \\
20  &  2228  &  Time zones  &  what time zone is in \textit{(905691)}  \\
21  &  2571  &  Opening hours \& events  &  what time does shoe carnival close \textit{(904553)}  \\
22  &  201  &  Hiking trails  &  how long is the trail to hike the narrows \textit{(265295)}  \\
23  &  300  &  Energy \& electricity  &  what energy source is used most by the usa \textit{(657321)}  \\
24  &  206  &  Latitude \& longitude  &  latitude is measured from where \textit{(438015)}  \\
25  &  301  &  Restaurants  &  what restaurant are open \textit{(891272)}  \\
26  &  307  &  University locations  &  which state is harvard university located \textit{(1019701)}  \\
27  &  449  &  Company ownership  &  who owns the ritz carlton chain \textit{(1045915)}  \\
28  &  1801  &  Hotels \& resorts  &  where is the resort that has the rooms over the water \textit{(997370)}  \\
29  &  476  &  Company headquarters  &  where is headquarters located \textit{(984270)}  \\
30  &  1797  &  Sports teams \& venues  &  where is the college football playoff \textit{(995090)}  \\
31  &  375  &  TV \& radio stations  &  what networks stream live tv \textit{(881802)}  \\
32  &  367  &  School calendars  &  what day does school start in florida \textit{(616616)}  \\
33  &  2368  &  Airports  &  what is closest airport \textit{(731434)}  \\
34  &  4494  &  Distances between places  &  how far is it between two cities \textit{(231276)}  \\
35  &  1082  &  Manufacturing locations  &  where does ford manufacture \textit{(973047)}  \\
36  &  469  &  Stores \& retail chains  &  how many shops does walmart have \textit{(295729)}  \\
37  &  526  &  Where to buy products  &  Where Can I Buy Cruex \textit{(8025)}  \\
38  &  249  &  Burial sites  &  name of where  dead bodies are buried \textit{(461873)}  \\
39  &  562  &  Shootings \& violent deaths  &  how many deaths from the shooting in fl \textit{(1097258)}  \\
40  &  321  &  University rankings \& accreditation  &  top ranked universities in us \textit{(522650)}  \\
41  &  247  &  Student enrollment numbers  &  how many students does osu have \textit{(297340)}  \\
42  &  391  &  University admissions / tests  &  average gpa acceptance at unr \textit{(36767)}  \\
43  &  250  &  Small towns \& forts  &  what part of maine is portland in \textit{(884763)}  \\
44  &  237  &  Surname \& place name origins  &  where does the name ireland come from \textit{(974631)}  \\
45  &  533  &  Cuisines by country  &  what are the names of two traditional foods in argentina \textit{(571955)}  \\
46  &  661  &  Origins of foods \& things  &  where do chickens come from originally \textit{(1141679)}  \\
47  &  390  &  Bridges \& tunnels  &  longest bridge in \textit{(1174000)}  \\
48  &  344  &  Tall buildings \& towers  &  what is the tallest building in the world \textit{(849724)}  \\
49  &  526  &  Islands  &  what island is off the map \textit{(865244)}  \\
\bottomrule
\end{tabular}
\caption{Resulting clusters with final labels and representative queries.}
\label{tab:cluster-rep-queries}
\end{table}

\begin{table}
\centering
\footnotesize
\begin{tabular}{r r p{5cm} p{7.5cm}}
\toprule
\textbf{\#} & \textbf{Size} & \textbf{Label} & \textbf{Representative query \textit{(query ID)}} \\
\midrule
50  &  368  &  Lakes \& natural water features  &  what is the largest freshwater lake in the united states \textit{(827162)}  \\
51  &  255  &  Native American tribes  &  where native american came from \textit{(1000891)}  \\
52  &  214  &  Pre-Columbian civilisations  &  what ancient civilizations were in mexico \textit{(553539)}  \\
53  &  396  &  Rivers  &  which river is the longest \textit{(1139538)}  \\
54  &  948  &  Mountains \& elevation  &  which mountain is the tallest \textit{(1013640)}  \\
55  &  1015  &  US places  &  what is the county of florida \textit{(813209)}  \\
56  &  480  &  Foreign governments \& leaders  &  what type of government does mexico have currently \textit{(912604)}  \\
57  &  328  &  Flags \& national symbols  &  what do the colors of the flag say about the nation \textit{(625172)}  \\
58  &  440  &  Video game locations  &  where do i find quarried stone skyrim \textit{(970929)}  \\
59  &  1046  &  Ancient civilisations  &  what is the ancient name of the area \textit{(804856)}  \\
60  &  1080  &  Health \& disease statistics  &  how many sick people in the united states \textit{(295798)}  \\
61  &  345  &  World population by country  &  what the world population \textit{(903847)}  \\
62  &  615  &  Oceans, water \& hydrology  &  most of the water on earth is found where \textit{(458406)}  \\
63  &  210  &  Glaciers, rifts, ocean depths  &  what is the deepest part of the ocean \textit{(814391)}  \\
64  &  2403  &  US local population statistics  &  what is the population of wv \textit{(1034203)}  \\
65  &  811  &  Construction dates of landmarks  &  what year was the statue of liberty constructed \textit{(928604)}  \\
66  &  291  &  Mining \& natural resources  &  where is silver ore found \textit{(992908)}  \\
67  &  245  &  Fish \& freshwater fishing  &  what fish are in lake nottely \textit{(1117541)}  \\
68  &  271  &  Vietnam War  &  when did the us go to vietnam war \textit{(942467)}  \\
69  &  1182  &  Languages spoken  &  what languages are spoken \textit{(872702)}  \\
70  &  909  &  Regions and countries  &  asia where is it located \textit{(27347)}  \\
71  &  314  &  Plate tectonics  &  when oceanic crust and continental crust meet at the plate boundary \textit{(954075)}  \\
72  &  1045  &  Planets \& space  &  what is the nearest planet to earth \textit{(836079)}  \\
73  &  330  &  Earth's layers and atmosphere  &  what layer of the earth is on the surface \textit{(872996)}  \\
74  &  561  &  Volcanoes  &  what type of eruption formed the volcano in this photograph \textit{(912074)}  \\
75  &  598  &  Rock types  &  granite is what type of rock \textit{(196556)}  \\
76  &  239  &  Germany \& France geography  &  germanys geographic location \textit{(194599)}  \\
77  &  979  &  Italy, Spain \& Mexico geography  &  what is italy's location \textit{(761674)}  \\
78  &  221  &  Pearl Harbor \& WWII Japan  &  when did the japanese attack pearl harbor \textit{(941810)}  \\
79  &  236  &  Constitution \& voting rights  &  how many states voted to ratify the constitution \textit{(296971)}  \\
80  &  236  &  Where plants \& fruits grow  &  where do pomegranates grow \textit{(971421)}  \\
81  &  842  &  Gardening \& plant zones  &  what planting zone am i \textit{(887313)}  \\
82  &  579  &  Biomes \& ecosystems  &  place where desert are found and their plants \textit{(475668)}  \\
83  &  1594  &  Animal habitats  &  where are most of the animals located \textit{(965819)}  \\
84  &  794  &  Founding of colonies \& countries  &  dates that colonies were founded quizlet \textit{(115607)}  \\
85  &  334  &  US statehood dates  &  what year did states become states \textit{(926788)}  \\
86  &  612  &  Battles \& military history  &  what was the last major battle of the civil war \textit{(921122)}  \\
87  &  514  &  Germany \& World Wars  &  what did the construction of the berlin wall do to germany \textit{(619964)}  \\
88  &  433  &  US places  &  in what county is washington wi \textit{(394062)}  \\
89  &  262  &  US places  &  what county in minneapolis mn \textit{(602024)}  \\
90  &  258  &  US places  &  lansing mi is in what county \textit{(435687)}  \\
91  &  1241  &  US places  &  austin texas  is in what county \textit{(29756)}  \\
92  &  474  &  US places  &  what county richmond va \textit{(615048)}  \\
93  &  205  &  US places  &  what county is atlanta \textit{(602628)}  \\
94  &  258  &  US places  &  what county is kansas city mo in \textit{(607886)}  \\
95  &  456  &  US places  &  what county is new york ny \textit{(610266)}  \\
96  &  295  &  US places  &  what county is cambridge ma \textit{(603814)}  \\
97  &  467  &  US places  &  in what county is columbus, oh \textit{(393920)}  \\
98  &  382  &  US places  &  what county is chicago in illinois \textit{(604215)}  \\
99  &  555  &  US places  &  what county in lebanon pa in \textit{(602014)}  \\
100  &  213  &  US places  &  what county is jersey city  nj \textit{(607788)}  \\
101  &  750  &  US places  &  what county is farmville nc in \textit{(605858)}  \\
102  &  342  &  US places  &  what county nashville tn in \textit{(615017)}  \\
103  &  869  &  US places  &  valencia is in what county in ca \textit{(1089047)}  \\
104  &  381  &  US places  &  what county portland in \textit{(615046)}  \\
\bottomrule
\end{tabular}
\caption*{\textbf{Table \ref{tab:cluster-rep-queries} (continued)}}
\label{tab:cluster-rep-queries-2}
\end{table}